\documentclass[aps,pra,twocolumn,superscriptaddress,showpacs]{revtex4-1}
\def\be{\begin{equation}}
\def\ee{\end{equation}}
\def\bea{\begin{eqnarray}}          
\def\eea{\end{eqnarray}}
\def\bi{\begin{itemize}}
\def\ei{\end{itemize}}

\usepackage{graphicx}
\usepackage{xcolor}

\begin{document}

\title{ 
                 Projected Entangled Pair States at Finite Temperature:  \\
                         Imaginary Time Evolution with Ancillas
}

\author{Piotr Czarnik} 
\affiliation{Instytut Fizyki Uniwersytetu Jagiello\'nskiego
             and Centre for Complex Systems Research,
             ul. Reymonta 4, 30-059 Krak\'ow, Poland}
             
\author{Lukasz Cincio} 
\affiliation{Perimeter Institute for Theoretical Physics, Waterloo, Ontario, N2L 2Y5, Canada}

\author{Jacek Dziarmaga} 
\affiliation{Instytut Fizyki Uniwersytetu Jagiello\'nskiego
             and Centre for Complex Systems Research,
             ul. Reymonta 4, 30-059 Krak\'ow, Poland}

\date{ September 1, 2012 }

\begin{abstract}
A projected entangled pair state (PEPS) with ancillas is evolved in imaginary time.
This tensor network represents a thermal state of a 2D lattice quantum system. 
A finite temperature phase diagram of the 2D quantum Ising model in a transverse field 
is obtained as a benchmark application.
\pacs{ 03.67.-a, 03.65.Ud, 02.70.-c, 05.30.Fk }
\end{abstract}

\maketitle

\section{ Introduction } 
Quantum tensor networks are a competitive tool to study strongly correlated quantum systems on a lattice. 
They originate from the density matrix renormalization group \cite{White} - an algorithm to minimize 
the energy of a matrix product state (MPS) ansatz in one dimension (1D). In recent 
years the MPS was generalized to a 2D ``tensor product state'' better known as a projected entangled pair
state (PEPS) \cite{PEPS}. Another type of tensor network is the multiscale entanglement renormalization ansatz 
(MERA) \cite{MERA} that is, in some respects, a refined version of the real space renormalization group. 
Both PEPS and MERA can be applied to strongly correlated fermions in 2D \cite{fermions}, because they do not 
suffer from the notorious fermionic sign problem. This makes them a powerfull tool to attack some of the hardest 
problems in strongly correlated electronic systems, including the enigmatic high temperature superconductivity 
\cite{highTc}. Indeed, PEPS has already provided first results for the ground state energy of the $t-J$ model 
\cite{tJ}, that can compete with the best variational Monte-Carlo results \cite{VMC}.

In contrast to the ground state, thermal states have been explored mainly with the MPS in 1D \cite{ancillas,WhiteT},
but they are more interesting in 2D, where they can undergo finite temperature phase transitions. In 2D thermal 
states were represented by tensor product states and contracted with the help of the higher-order singular 
value decomposition in Ref. \cite{HOSVD}. A similar projected entangled-pair operator (PEPO) ansatz was proposed in 
Ref. \cite{PEPO}. 

In this paper we follow a different route. In a way that can be easily generalized to 2D, the MPS can be extended 
to finite temperature by appending each lattice site with an ancilla \cite{ancillas}. A thermal state is obtained by 
imaginary time evolution of a pure state in the enlarged Hilbert space, starting from infinite temperature. Unfortunately, 
in contrast to 1D, where the time evolution of a MPS can be simulated accurately and efficiently, in 2D the time evolution 
of PEPS appears to be a hard problem. It requires accurate computation of a tensor environment that is often hard to 
approximate accurately and reliably. The aim of this work is to overcome this problem.     

\section{ Thermal states } 
We consider spins on an infinite square lattice with a Hamiltonian ${\cal H}$. 
Every spin has $S$ states $i=0,...,S-1$ and is accompanied by an ancilla with states $a=0,...,S-1$. 
The enlarged Hilbert space is spanned by states $\prod_s |i_s,a_s\rangle$, where the product runs over lattice sites $s$. 
The state of spins at infinite temperature,
$
\rho(\beta=0) = \prod_s \left( \frac{1}{S} \sum_{i=0}^{S-1} | i_s \rangle\langle i_s| \right) \propto {\bf 1},
$
is obtained from a pure state in the enlarged Hilbert space, 
\be 
\rho(0) ~=~ {\rm Tr}_{\rm ancillas}|\psi(0)\rangle\langle\psi(0)|~,
\ee
where 
\be 
|\psi(0)\rangle ~=~ \prod_s \left( \sum_{i=0}^{S-1} \frac{1}{\sqrt{S}} |i_s,i_a\rangle \right)~
\label{psi0}
\ee
is a product of maximally entangled states of every spin with its ancilla. The state $\rho(\beta)\propto e^{-\beta {\cal H}}$ 
at finite $\beta$ is obtained from
\be 
|\psi(\beta)\rangle~\propto~
e^{-\frac12\beta {\cal H}}~|\psi(0)\rangle~\equiv~
U(\beta)~|\psi(0)\rangle~
\ee
after imaginary time evolution for time $\beta$ with $\frac12{\cal H}$.

\begin{figure}[t!]
\includegraphics[width=1.0\columnwidth,clip=true]{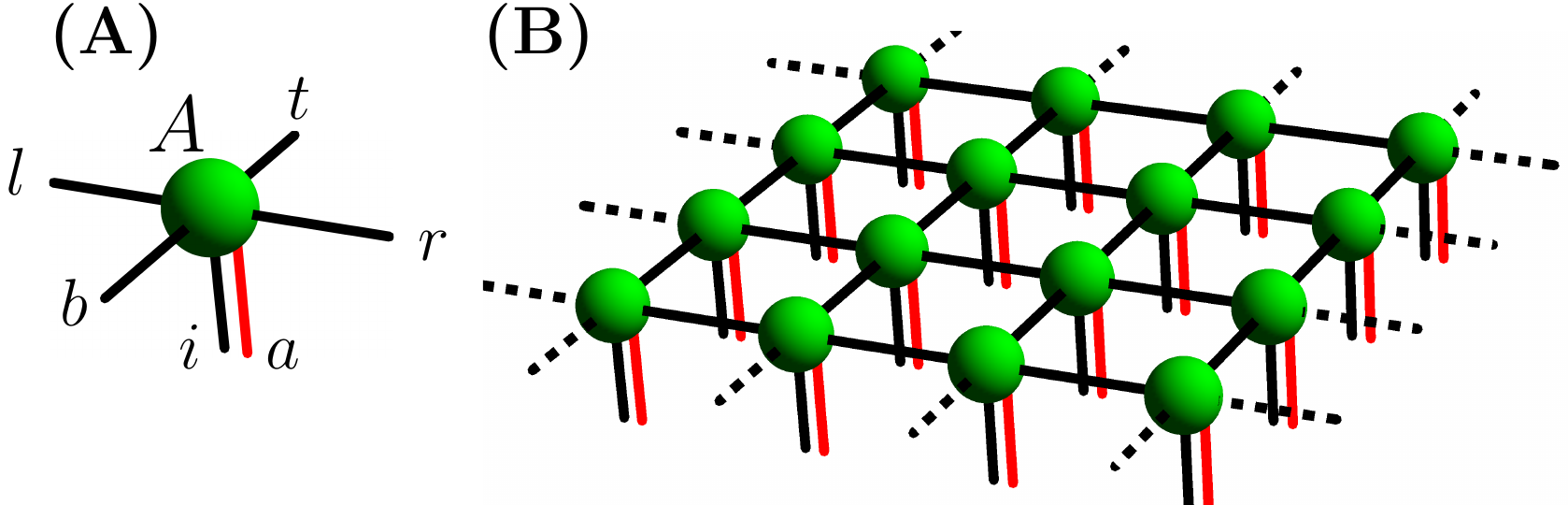}
\caption{ 
In A, graphic representation of the tensor $A^{ia}_{trbl}$.
In B, the amplitude $\Psi_A[\{i,a\}]$ with all bond indices connecting nearest-neighbor tensors contracted.
The index contraction is represented by a line connecting two tensors.
}
\label{FigPeps}
\end{figure}

\section{ PEPS } 
In the quantum Ising model with spin-$\frac12$ that we consider in the rest of this paper, the translational invariance is not broken and a unit cell encloses only one lattice site. 
Therefore, for an efficient simulation of the time evolution we represent $|\psi(\beta)\rangle$ by a translationally invariant PEPS with the same tensor $A^{ia}_{trbl}(\beta)$ at every site. 
Here $i,a = 0,...,S-1$ are the spin and ancilla indices respectively, $S=2$, 
and $t,r,b,l=0,...,D-1$ are the bond indices to contract the tensor with similar tensors at the nearest neighbor sites, 
see Fig. \ref{FigPeps}A. 
The ansatz is
\be 
|\psi(\beta)\rangle ~=~  \sum_{\{i_s,a_s\}} ~ \Psi_A[\{i_s,a_s\}] ~ \prod_s|i_s,a_s\rangle~ \equiv~ \left|\psi_A\right\rangle.
\ee
Here the sum runs over all indices $i_s,a_s$ at all sites $s$. 
The amplitude $\Psi_A$ is the tensor contraction in Fig. \ref{FigPeps}B. 
The initial state (\ref{psi0}) can be represented by a tensor 
\be 
A^{ia}_{trbl}~=~ \delta^{ia} ~ \delta_{t0} ~ \delta_{r0} ~ \delta_{b0} ~ \delta_{l0}~.
\label{A0}
\ee
$D=1$ is the minimal bond dimension sufficient to represent the initial state.

\section{ Quantum Ising model in 2D } 

We proceed with
\be 
{\cal H} ~=~- \sum_{\langle s,s'\rangle}Z_sZ_{s'}-h \sum_s X_s ~\equiv~ {\cal H}_{ZZ}+{\cal H}_X.
\label{calH}
\ee
Here $Z,X$ are Pauli matrices. 
The model has a ferromagnetic phase with non-zero spontaneous magnetization $\langle Z \rangle$ for small $h$ and large $\beta$. 
At $h=0$ the critical point is $\beta_c=-\ln(\sqrt{2}-1)/2=0.441$, 
and at $\beta^{-1}=0$ the quantum critical point is $h_c=3.04$, see Ref. \cite{hc}.

\section{ Suzuki-Trotter decomposition }
  
We define 
$ 
U_{ZZ}(\Delta\beta) \equiv e^{-\frac12{\cal H}_{ZZ}\Delta\beta}
$ 
and 
$
U_X(\Delta\beta) \equiv e^{-\frac12{\cal H}_X\Delta\beta}
$ 
for the interaction and the transverse field respectively. In the Suzuki-Trotter decomposition a small time step 
\be
U(d\beta) ~=~ U_X(d\beta/2)U_{ZZ}(d\beta)U_X(d\beta/2)~+~{\cal O}(d\beta^3).
\ee
The action of $U_X(d\beta)$ on PEPS replaces $A^{ia}_{trbl}$ with
\be 
\propto ~A^{ia}_{trbl}~+~\epsilon~\sum_{j=0,1}X^{ij}A^{ja}_{trbl} 
\ee
of the same bond dimension $D$. Here $\epsilon=\tanh\left(\frac12h~d\beta\right)$. However, the action of $U_{ZZ}(d\beta)$ 
maps $A$ to a new tensor
\be 
B^{ia}_{2t+s_t,2r+s_r,2b+s_b,2l+s_l} ~\propto~ \epsilon^{s/2}~(-1)^{is}~A^{ia}_{trbl}~. \label{B}
\ee
Here $\epsilon=\tanh\left(\frac12d\beta\right)$, indices $s_t,s_r,s_b,s_l\in\left\{0,1\right\}$, and $s=s_t+s_r+s_b+s_l$. 
This is an exact map, but $B$ has the bond dimension $2D$ instead of the original $D$. 

\section{ Tensor renormalization }  
\begin{figure}[t!]
\includegraphics[width=1.0\columnwidth,clip=true]{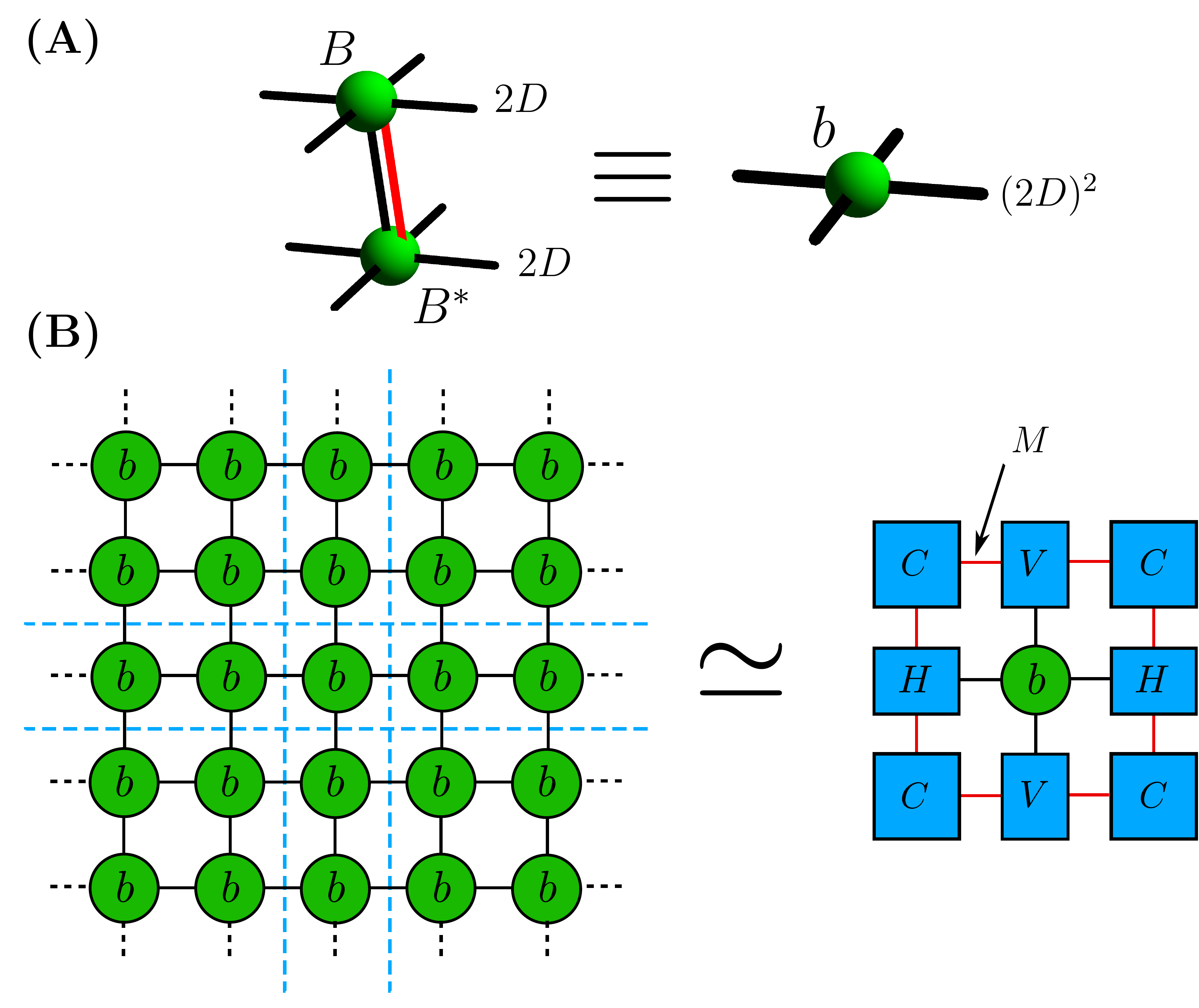}
\caption{ 
In A, the contraction of two tensors $B$ on the left hand side (LHS) gives the transfer matrix $b$ on the right hand side (RHS), as seen from the top.
In B, the contraction of transfer matrices on the LHS is the norm squared ${\rm Tr}~\rho(\beta)=\langle\psi(\beta)|\psi(\beta)\rangle$.
This contraction cannot be done exactly. It is approximated by the contraction on the RHS with a corner matrix $C$ and vertical/horizontal tensors $V,H$. 
Their (red) environmental indices have dimension $M$. The $C,V,H$ are such that to the transfer matrix $b$ in the center its environment on the RHS should
appear the same as its exact environment on the LHS. Their construction is described in Fig. \ref{FigRenC}.
}
\label{FigCVH}
\end{figure}

\begin{figure}[t!]
\includegraphics[width=0.7\columnwidth,clip=true]{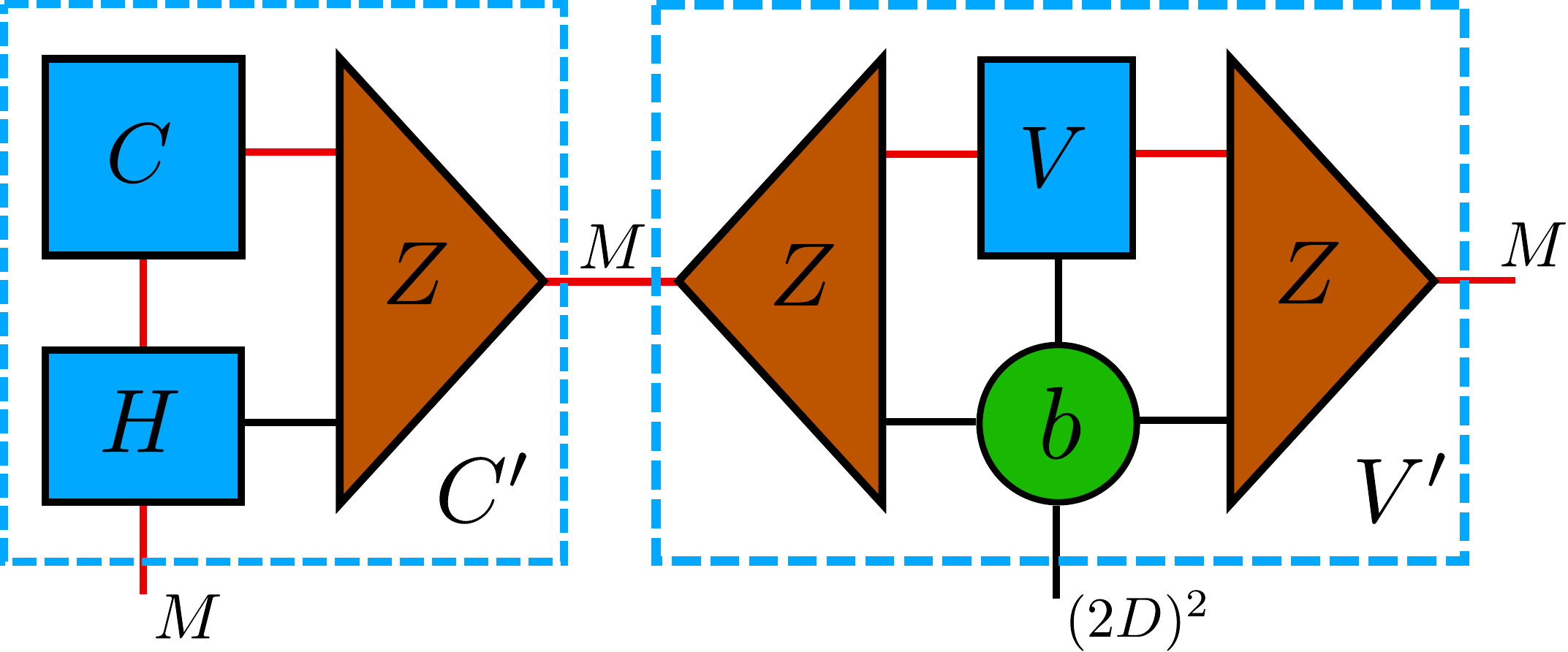}
\caption{ 
The tensors $C,V,H$ are obtained by repeating a renormalization procedure until convergence. The procedure has two steps. In the first step
the tensors $C$ and $H$ are contracted to form a matrix $C.H$. The $M\times M(2D)^2$ matrix $C.H$ is subject to singular value decomposition. It has $M$ right 
singular vectors that define an isometry $Z$. The isometry is used to compress the right index of $C.H$ back to dimension $M$ giving a new corner matrix $C'$. 
The same $Z$ renormalizes the contraction $V.b$ giving a new $V'$. The second step is the same but with the roles of $H$ and $V$ interchanged. The two-step 
procedure is repeated until convergence measured by the figure of merit explained in Fig. \ref{FigRenB}.
}
\label{FigRenC}
\end{figure}

\begin{figure}[t!]
\includegraphics[width=0.99\columnwidth,clip=true]{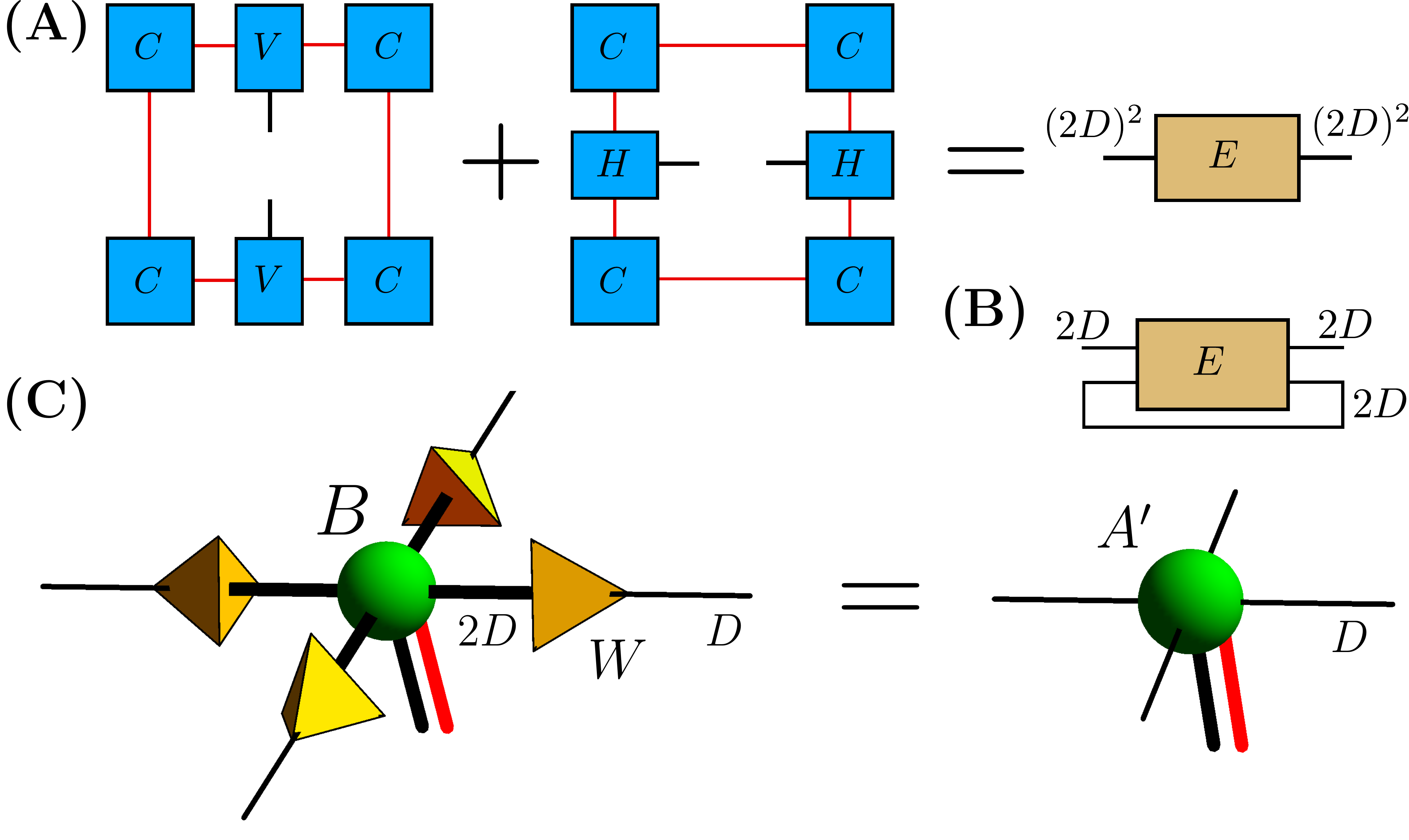}
\caption{ 
In A, 
for the isotropic tensors $A$ 
the diagrams on the LHS would be two equivalent representations of the norm squared of the state, 
if not for the one uncontracted bond in the middle of each of them. 
For better numerical stability, 
we add these equivalent diagrams making a square matrix $E$ of dimension $(2D)^2$ on the RHS. 
By construction, $E$ is non-negative and its trace is equal to the norm squared. 
(Renormalization of a complex and non-symmetric E is described in Appendix \ref{AppA}.)
${\rm Tr}~E$ is the figure of merit refered to in the caption of Fig. \ref{FigRenC}.
In B, 
each of the two indices of $E$ can be represented by two indices of dimension $2D$ in such a way that the upper(lower) index corresponds to the top(bottom) layer of tensors $B$ in Fig. \ref{FigCVH}A. After the lower index is traced out we obtain a non-negative $2D\times2D$ matrix. Its $D$ leading eigenvectors corresponding to the $D$ largest eigenvalues
define the isometry $W$. 
In C, 
the isometry renormalizes the new tensor $B$ back to a tensor $A'$ with the original bond dimension $D$ thus completing the action of $U_{ZZ}$ on PEPS.
}
\label{FigRenB}
\end{figure}

The bond dimension has to be truncated back to $D$ in a way least distortive to the new PEPS $|\psi_B\rangle$. 
The general idea is to use an isometry $W$ that maps from $2D$ to $D$ dimensions:
\be 
\sum_{t',r',b',l'=0}^{2D-1}
W_t^{t'}~
W_r^{r'}~
W_b^{b'}~
W_l^{l'}~
B^{ia}_{t'r'b'l'}~=~
A'^{~ia}_{trbl}~,
\ee 
see also Fig. \ref{FigRenB}C. The isometry should be the least destructive to the norm squared $\langle\psi_B|\psi_B\rangle$. 
The construction of the optimal isometry described in Figs. \ref{FigCVH},\ref{FigRenC},\ref{FigRenB} is a variant
of the corner matrix renormalization \cite{CMR}.
It requires calculation of a tensor environment of $B$ in the network representing $\langle\psi_B|\psi_B\rangle$.
Unfortunately, this environment cannot be calculated exactly in an efficient way. 
This is why it is replaced by an effective environment, made of environmental tensors $C,V,H$,
that should appear to the tensor $B$ the same as the exact one as much as possible.
The environmental tensors are contracted with each other by indices of dimension $M$.   
Increasing $M$ should make the effective environment more accurate. 
The overall cost of renormalizing $B$ back to the bond dimension $D$ is polynomial in both $D$ and $M$. 
It is dominated by the calculation of $V'$ in Fig. \ref{FigRenC} that scales like 
$M^3D^4$ when $M\geq D^4$ or $M^2D^8$ otherwise. 


At the beginning of the evolution the environmental tensors $C,V,H$ are initialized with random numbers and, 
in principle, they should be reinitialized after every time step. This, however, would not be the most efficient method for
a smooth time evolution where both $A$ and the environmental tensors change infinitesimally 
in an infinitesimal time step. Thus after every time step it might be more efficient to use the converged environmental tensors 
as the initial ones for the next step. This ``recycling'' would accelerate convergence in the 
next step because there would be very little to converge. However, we found that such ``recycled''
evolution is very fast indeed but, especially near a phase transition, the tensors get trapped in lower dimensional 
subspaces, not making full use of the available dimension $M$. Results often appear converged in increasing $M$ while
they are actually just trapped in an $M_{\rm eff}<M$. This is not quite surprising, because even though the tensor $A$ 
may evolve smoothly across a critical point, the environment does not need to be smooth, 
because it represents the rest of the infinite system at criticality.  
It is the environment that is critical, not $A$, even though the environment is made of an infinite number of smooth $A$'s. 
To prevent the trapping, but at the same time not to slow down the algorithm too much, we add weak noise to 
the converged tensors before they are reused in the next time step. In practice, a noise at the level of $<1\%$ of a typical 
tensor element was enough for the algorithm to produce the same results as if the tensors were reinitialized, but at a much faster rate. 
Since we want an accurate time evolution, it is essential that the tensors do not get 
trapped in any single time step, because the errors can accumulate and derail the following evolution.

Another technical issue concerns the construction of $V'$ in Fig. \ref{FigRenC}. In principle, all singular vectors $Z$ of the corner matrix should
be used in this contraction, even those corresponding to singular values equal to numerical zero. The ``zero vectors'' 
do not make any difference when $V'$ is contracted with $C'$. However, we found the algorithm unstable unless we set the zero 
vectors in $V'$ to zero. These (numerically inaccurate) vectors do make a difference when $V'$ is contracted with a tensor other
than $C'$ and this opens room for the observed instability.

\section{ Zero transverse field } 
In this classical limit the exact state $|\psi(\beta)\rangle=U_{ZZ}(\beta)|\psi(0)\rangle$ can be obtained from the initial state
by just one application of $U_{ZZ}(\beta)$. As in Eq. (\ref{B}), this exact transformation doubles the bond dimension of the 
initial tensor (\ref{A0}) to $D=2$. Thus $D=2$ (or $D=S$ in general) is enough for an exact PEPS representation of any classical state including 
the critical one. However, calculation of expectation values requires an approximate environment build with the tensors $C,V,H$ of 
a finite dimension $M$. The closer to criticality the bigger $M$ is needed to represent the critical correlations in the environment. 

\begin{figure}[t!]
\includegraphics[width=1.0\columnwidth,clip=true]{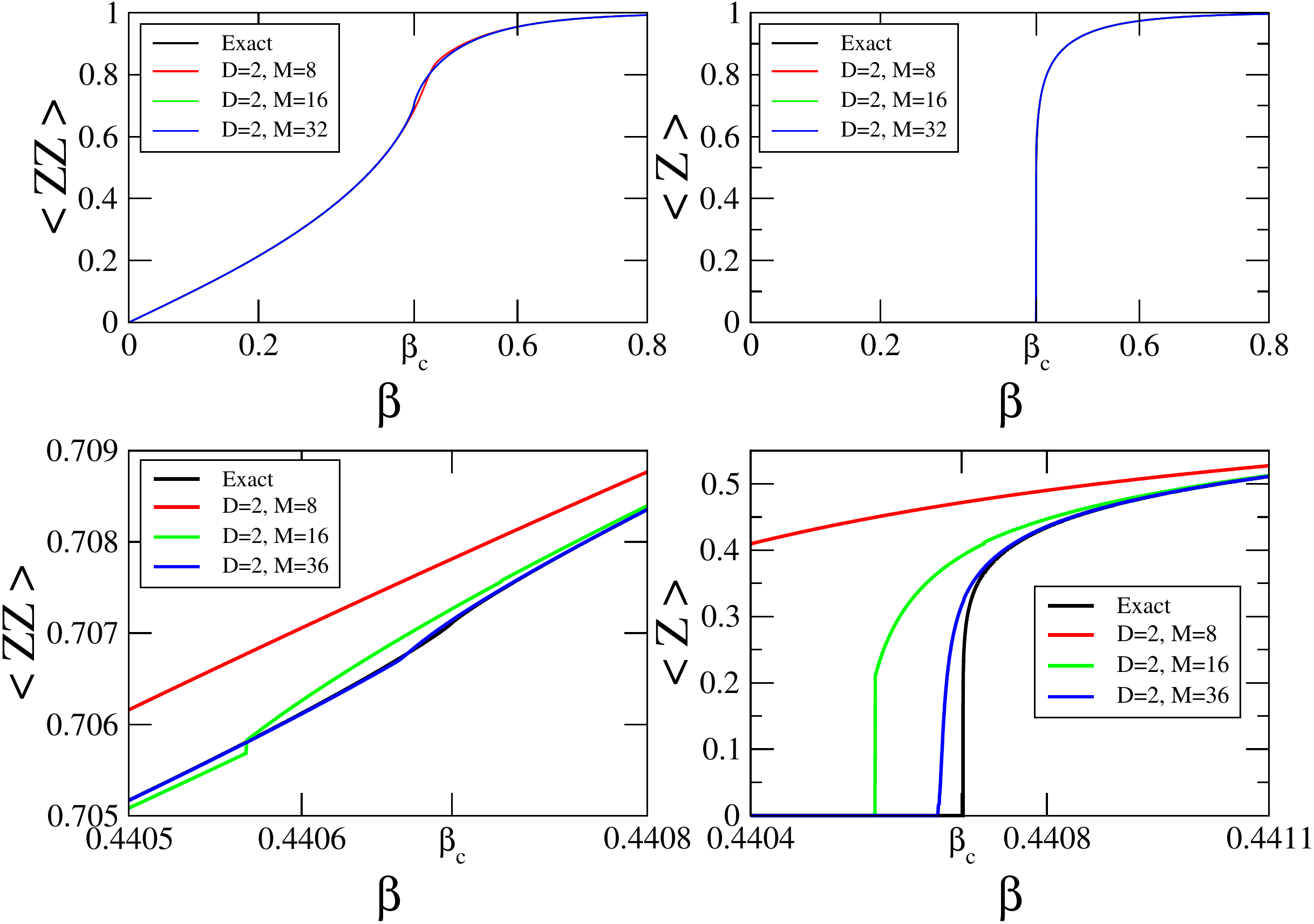}
\caption{ 
Numerical results versus exact Onsager's solution in the classical case of zero transverse field $h=0$. 
Here $\langle ZZ\rangle$ is the ferromagnetic correlator between nearest neighbors, and $\langle Z\rangle$
is spontaneous magnetization. The upper panels show these quantities in a wide range of 
$\beta$, and the lower panels zoom within $10^{-3}$ of the critical $\beta_c$. PEPS with $D=2$ is 
an exact representation of a thermal state, but accurate evolution and calculation of 
expectation values near $\beta_c$ require large $M$. The lower panels show their 
convergence to the exact solution with increasing $M$. Here $d\beta=10^{-4}\beta_c$ 
(upper panels) and $d\beta=10^{-6}\beta_c$ (lower panels).
}
\label{FigOnsager}
\end{figure}

Figure \ref{FigOnsager} shows numerical simulations of the evolution by a product of small transformations
$U_{ZZ}(d\beta)$. After each transformation the tensor is renormalized back to $D=2$. The plots show excellent agreement with
Onsager's solution, except in a narrow neighborhood of the critical point, but even there increasing $M$ seems to converge the numerical solution towards the exact one. 

The numerical results suggest that at the critical point a very large, if not infinite, $M$ is needed for an accurate description of the long range critical correlations. Consequently, imaginary time evolution with a finite-$M$ is bound to accumulate unrecoverable errors near the critical point that will distort the following low temperature phase. To avoid this distortion, we suggest to add a tiny symmetry-breaking term to the Hamiltonian in order to smooth out the non-analyticity of the critical point and turn it into a smooth crossover. A PEPS with a finite-$M$ can be evolved accurately across the crossover and into the low temperature phase. Eventualy the criticality can be recovered by turning the symmetry-breaking term down to zero and increasing $M$ at the same time. This is what we do below in the quantum case of a finite transverse field.

The spontaneous symmetry breaking in Fig. \ref{FigOnsager} may deserve a comment. In Eq. (\ref{B}) the zero temperature ferromagnetic state 
$U_{ZZ}(\infty)|\psi(0)\rangle$ is represented exactly by 
$
B^{ia}_{s_t,s_r,s_b,s_l} \propto (-1)^{is} ~ \delta^{ia}
$ 
that does not break the symmetry. Its transfer matrix is 
$
b_{S_t,S_r,S_b,S_l}
\propto
I_+^{S_t}I_+^{S_r}I_+^{S_b}I_+^{S_l}+I_-^{S_t}I_-^{S_r}I_-^{S_b}I_-^{S_l},
$
where $I_\pm=(1,\pm1)^T$ is a vector, $I_{\pm}^S$ is the $S$-th component of the vector $I_{\pm}$, 
$S_t=\left.s_t+\bar s_t\right|{\rm mod}~2$, and $s$'s($\bar s$'s) are the bond indices of the top(bottom) $B$ in Fig. \ref{FigCVH}A. 
The $I_\pm$-part of $b$ corresponds to $\langle Z \rangle=\pm 1$, 
but the symmetry between these two parts is broken by the iterative construction of the environmental tensors. 
Indeed, for $M=1$ the iterative procedure has two stable fixed points: 
$C_{11}=1,V_{1,S,1}=H_{1,S,1}=I^S_\pm$ breaking the symmetry to $\langle Z \rangle=\pm 1$. 
The same is true for $M=2$ when the stable symmetry-breaking points are $C_{S_1,S_2}=I^{S_1}_\pm I^{S_2}_\pm,V_{S_1,S_2,S_3}=H_{S_1,S_2,S_3}=I^{S_1}_\pm I^{S_2}_\pm I^{S_3}_\pm$.
By a suitable change of basis, each of these two fixed points can be represented by more compact tensors with $M=1$. Thus the symmetry breaking reduces the required $M$ from $2$ to $1$.
Once the symmetry is broken to a fixed point of the environment, the tensor renormalization in Fig. \ref{FigRenB}C also breaks the symmetry of the new renormalized tensor $A'$.
This simple example explains the property of the algorithm observed in the ferromagnetic phase: 
the symmetric state is unstable but, once the symmetry is broken, the broken state is more accurate than the symmetric one for the same $M$. 
The broken state is simply less entangled than the symmetric one.

\begin{figure}[t!]
\includegraphics[width=1.0\columnwidth,clip=true]{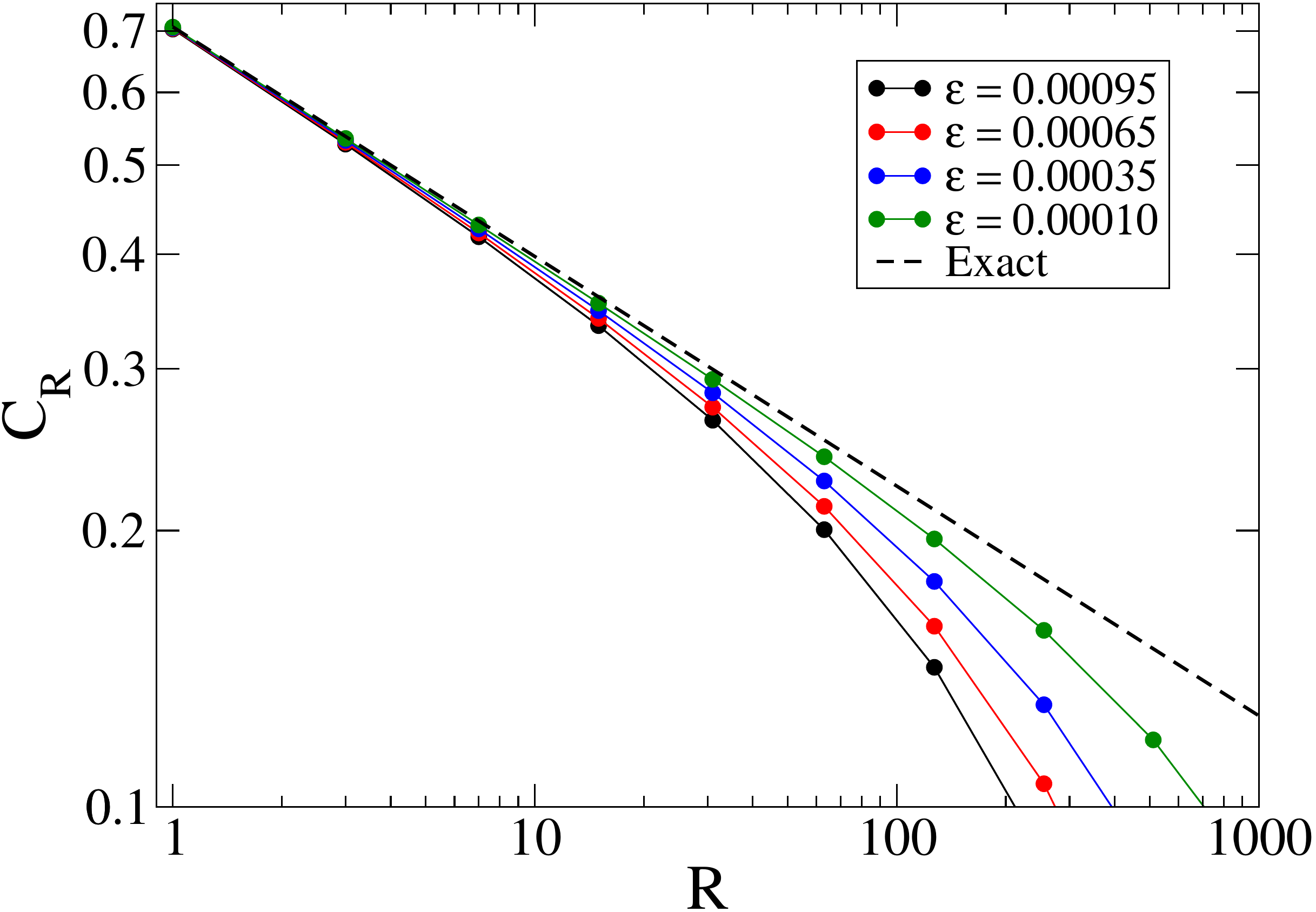}
\caption{ 
The correlator $C_R=\langle Z_{s} Z_{s'} \rangle$ as a function of the separation $R=|s-s'|$ 
at different distances from the critical point measured by $\varepsilon=(\beta_c-\beta)/\beta_c$.
The log-log plot shows convergence to the exact function $C_R\sim R^{-1/4}$ (dashed line). 
Here $D=2,M=32$, and $d\beta=10^{-6}\beta_c$.
}
\label{FigCR}
\end{figure}

\section{ Finite transverse field } 
\begin{figure}[t!]
\includegraphics[width=0.9\columnwidth,clip=true]{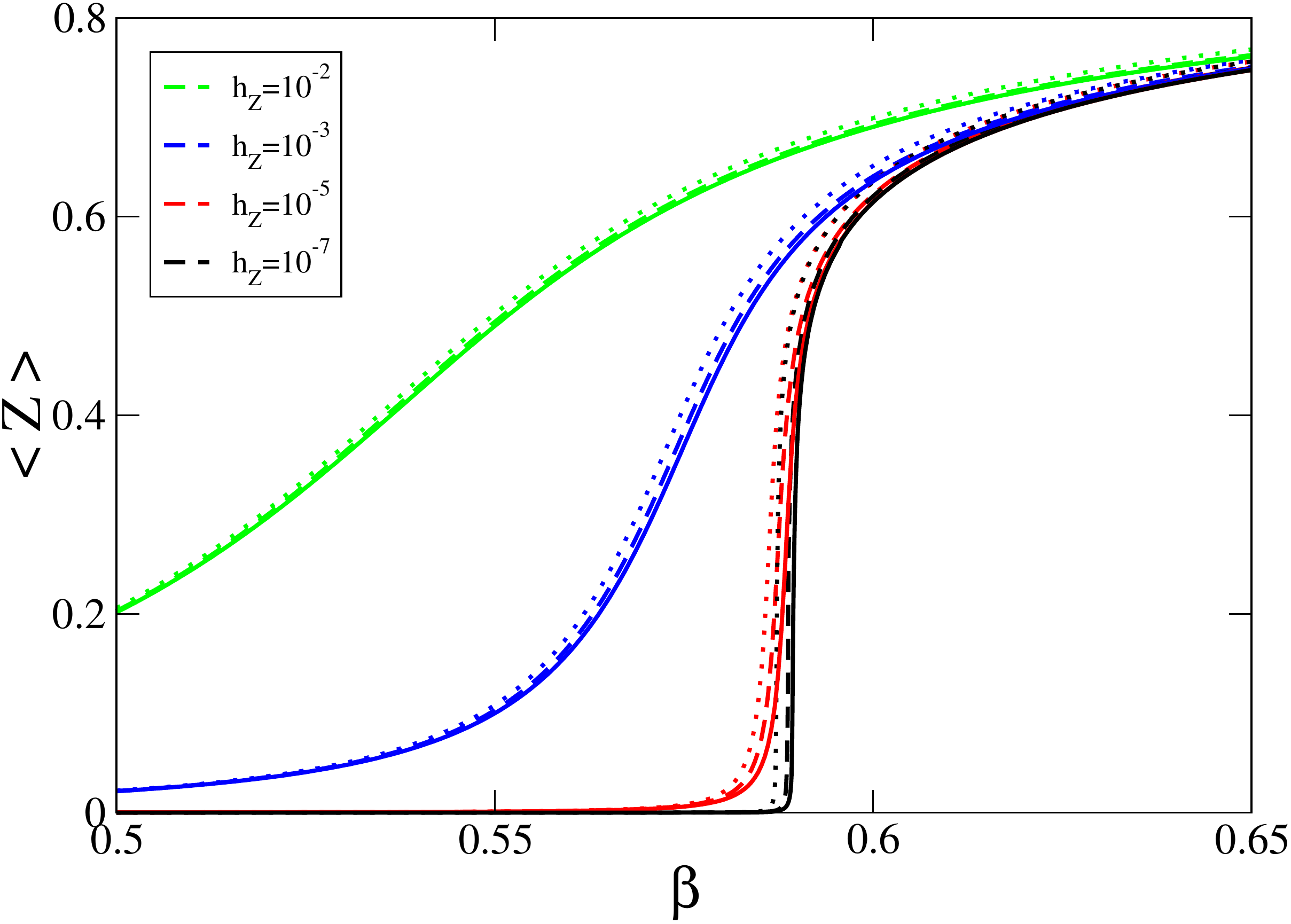}
\caption{ 
The magnetization $\langle Z\rangle$ as a function of $\beta$ for the quantum Hamiltonian with $h=\frac23h_c$ 
at different symmetry-breaking fields $h_Z\to0$. 
For $D=6$ (solid lines), $D=4$ (dashed lines), and $D=2$ (dotted lines) the magnetization converges to a non-analytic critical curve
when $h_Z\to0$. 
All plots are converged in $M$. The required $M\leq16$ increases with decreasing $h_Z$.
An adaptive imaginary time-step $d\beta$ was used with the shortest $d\beta\geq 10^{-6}\beta_c$ near the critical point.  
}
\label{cross}
\end{figure}

For $h>0$ the Hamiltonian (\ref{calH}) is quantum and a PEPS with a finite $D$ is in general not an exact representation
but an approximation to a thermal state. However, at finite temperature the fixed point of the renormalization group is 
a classical Hamiltonian whose critical thermal state can be represented by a PEPS exactly. This is why we expect that even 
a PEPS with the minimal non-trivial $D=2$ ($D=S$ in general) can in principle capture the universal critical properties of 
a quantum system at finite temperature, even though it may be not an accurate description of its short range quantum 
correlations. Just as in the classical case, it is mainly $M$ and not $D$ that limits the accuracy of PEPS at the critical 
point.

In order to smooth out the finite-$M$ imaginary time evolution across a critical point we added a small symmetry-breaking 
perturbation $\delta {\cal H}=-h_Z\sum_sZ_s$ with a tiny longitudinal field $h_Z$. The perturbation is rounding the 
non-analyticity at the critical point making it possible to evolve across the point with a finite $M$ without accumulating 
unrecoverable errors. We expect that accurate evolution will require increasing $M$ as $h_Z$ is turned down to $0$.  

\begin{figure}[t!]
\includegraphics[width=1.0\columnwidth,clip=true]{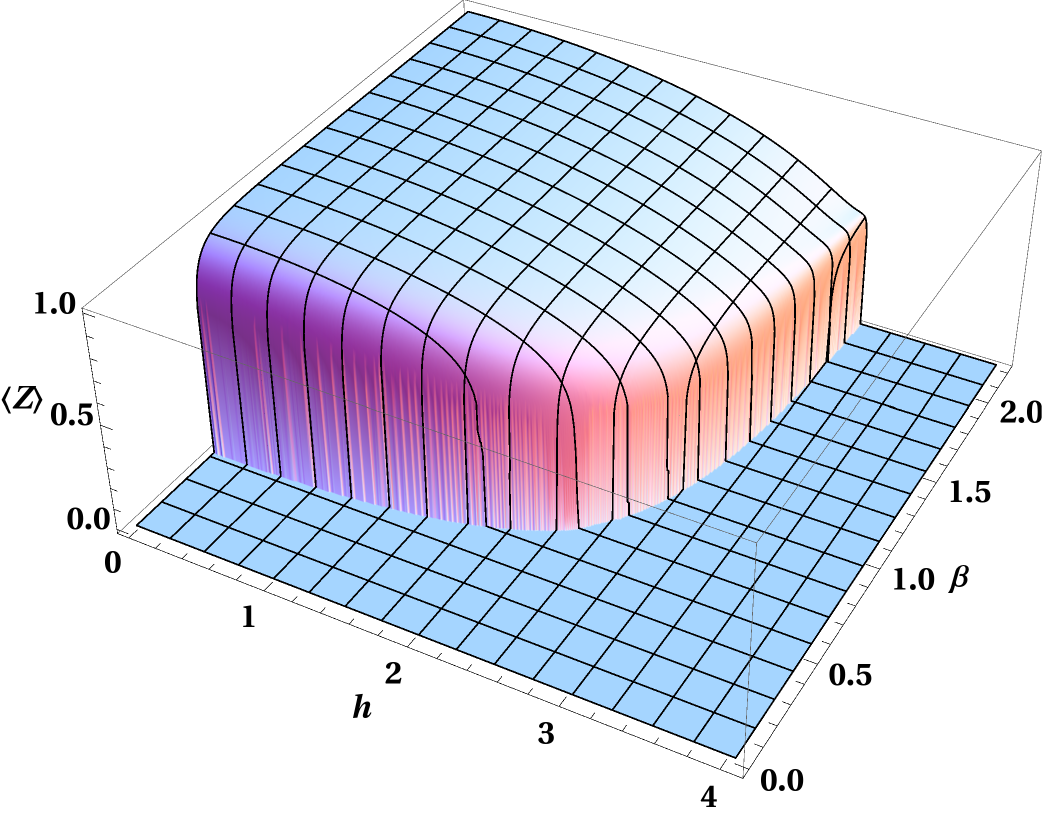}
\vspace{-1.0cm}
\caption{ 
The magnetization $\langle Z \rangle$ as a function of the transverse magnetic field $h$ 
and the inverse temperature $\beta$. This 3D plot clearly shows the ferromagnetic phase at low temperature
and weak transverse field.
Here $D=2$, $M=24$, $h_Z=10^{-10}$, and $d\beta=10^{-2}\beta_c$. 
}
\label{magnetization3D}
\end{figure}

Figure \ref{cross} shows numerical results for the magnetization $\langle Z\rangle$ at a relatively strong transverse 
field $h=\frac23h_c$. There is not much quantitative difference between the three sets of plots with $D=2,4,6$. As expected,
all three sets, even the minimal non-trivial $D=2$, converge to a non-analytic critical curve when $h_Z\to 0$. Encouraged by the 
cross-section in Fig. \ref{cross}, we also made a dense scan of the whole $h-\beta$ phase diagram with the minimal $D=2$.
The ferromagnetic phase at low temperature and weak transverse field can be clearly read from the 3D plot in Fig. \ref{magnetization3D}.

\section{ Conclusion } 
A PEPS with ancillas can be efficiently evolved in imaginary time generating a PEPS representation of thermal states.
In the case of a classical system, the bond dimension $D$ equal to the number of states per site is enough for an exact 
representation of any thermal state. The evolution was simulated with the Suzuki-Trotter decomposition accurate to the
second order in the time step. A variant of the corner matrix renormalization was used to obtain an accurate tensor environment.
After every time step, the environmental tensors were perturbed by a weak noise to ensure that they make full use of their dimensionality.

With some modification the algorithm can also evolve pure and thermal states in real time and, after introducing fermionic swap gates, generate finite temperature phase diagrams of strongly correlated fermions on a lattice \cite{future}.

\acknowledgements
We thank Guifr\'e Vidal for discussions, and Marek Rams for comments on the manuscript.
Work supported in part by the Polish National Science Center (NCN) grant 2011/01/B/ST3/00512 (P.C. and J.D.).

\appendix
\section{renormalization of general complex and non-symmetric PEPS tensors}\label{AppA}
\begin{figure}[t!]
\includegraphics[width=1.0\columnwidth,clip=true]{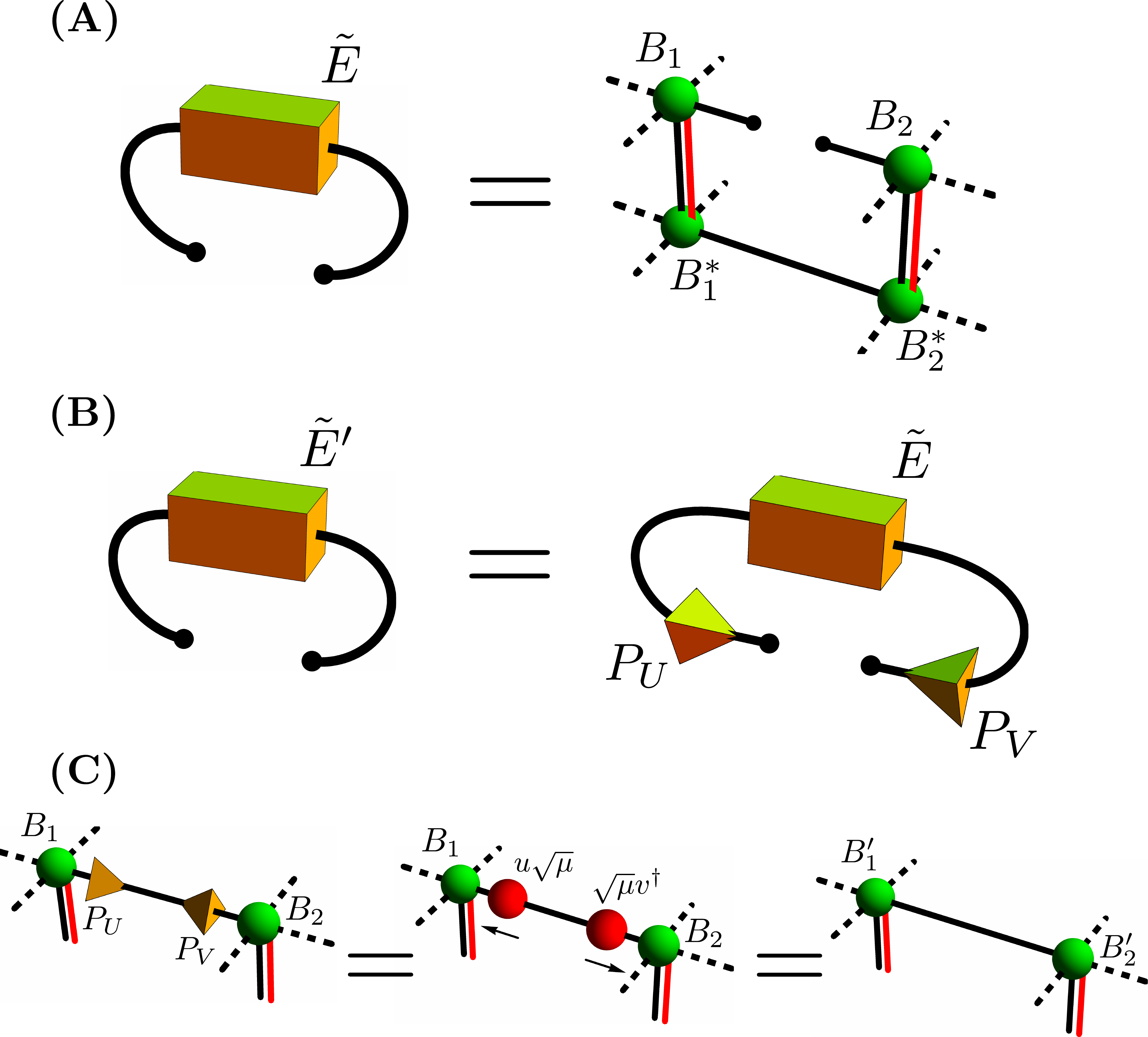}
\caption{ 
In A, 
the matrix $\tilde E$ arises from the tensor network representing the norm-squared $\langle\psi_B|\psi_B\rangle$ after cutting the bond between the nearest-neighbor PEPS tensors $B_1$ and $B_2$.
In B, 
the indices of $\tilde E$ are renormalized by the projectors $P_U$ and $P_V$.
In C,
since the indices of $\tilde E$ are also indices of the PEPS tensors $B_1$ and $B_2$, the PEPS tensors' indices are also renormalized by the projectors $P_U$ and $P_V$. The product 
$
P_UP_V=\sum_{\alpha,\beta=1}^{D}|U_\alpha\rangle~\langle U_\alpha|V_\beta\rangle~\langle V_\beta|
$ 
inserted in the bond between $B_1$ and $B_2$ results in a bond matrix $S_{\alpha\beta}=\langle U_\alpha|V_\beta\rangle$
on the bond $B_1-B_2$. The singular value decomposition of the bond matrix $S=u~\mu~v^\dag$ followed by absorbtion
of the matrix $u~\sqrt{\mu}$ into the right index of $B_1$ and the matrix $\sqrt{\mu}~v^\dagger$ into the left index of $B_2$
completes renormalization of the bond $B_1-B_2$. After renormalization of all bonds a new PEPS is obtained with
new tensors $A'$.  
}
\label{FigApp}
\end{figure}

The matrix in Fig. \ref{FigRenB}B, we will call it $\tilde{E}$ here, is used for the renormalization of the PEPS 
tensors from $B$ to $A'$. Its trace is the norm-squared of the PEPS: ${\rm Tr}\tilde E=\langle\psi_B|\psi_B\rangle$.
The matrix itself is the norm-squared in Fig. \ref{FigCVH}B, but with one of the bonds connecting pairs of nearest-neighbor
tensors $B$ cut open. For the isotropic tensors considered in this paper $\tilde E$ is by construction real, symmetric, and 
positive semi-definite. However, in general a PEPS may be neither isotropic, nor translationally invariant, nor even real,
and the matrix $\tilde E_{ij}$ is just a complex matrix. We show such a general matrix $\tilde E$ for a bond between 
inequivalent PEPS tensors $B_1$ and $B_2$ in Fig. \ref{FigApp}A. We want to renormalize the indices of this $\tilde E$, 
because they are also the indices of the PEPS tensors $B_1$ and $B_2$ that have to be renormalized back to the original 
bond dimension $D$. 

The renormalization procedure begins by a singular value decomposition
\be 
\tilde E~=~\sum_{\alpha=1}^{2D} |U_\alpha\rangle \lambda_\alpha \langle V_\alpha|~.
\label{tildeE}
\ee
Here $\lambda$'s are the singular values in decreasing order $\lambda_1\geq\lambda_2\geq...$, 
and $|U_\alpha\rangle$ ($|V_\alpha\rangle$) are corresponding left (right) singular vectors.
We define projectors $P_U=\sum_{\alpha=1}^{D}|U_\alpha\rangle\langle U_\alpha|$ and $P_V=\sum_{\alpha=1}^{D}|V_\alpha\rangle\langle V_\alpha|$.
A renormalized matrix is 
\be 
\tilde E'~=~P_U\tilde E P_V~=~\sum_{\alpha = 1}^{D} |U_\alpha\rangle \lambda_\alpha \langle V_\alpha| ~,
\ee
where we truncate the sum (\ref{tildeE}) from $2D$ to $D$, see Fig. \ref{FigApp}. This truncation minimizes the difference between $\tilde E$ and the renormalized $\tilde E'$ measured by the error
\be 
{\rm Tr} (\tilde E-\tilde E')^\dag(\tilde E-\tilde E') ~=~ \sum_{\alpha=D+1}^{2D} \lambda_\alpha^2 ~.
\ee
As mentioned above and shown in Fig. \ref{FigApp}A, the indices of $\tilde E$ that are renormalized by the projectors $P_U$ and $P_V$ are at the same time the bond indices of the nearest-neighbor PEPS tensors $B_1$ and $B_2$, see Fig. \ref{FigApp}C. 

The expression ${\rm Tr}\tilde E'$ is the norm-squared of the original PEPS, but with an additional ``bond matrix'' $S$ inserted in the bond connecting the renormalized nearest-neighbor PEPS tensors: 
\be 
S_{\alpha\beta}~=~\langle U_\alpha|V_\beta \rangle~.
\ee
Here $\alpha,\beta=1,...,D$. In order to go back to the original PEPS structure, without any bond matrices, we want to absorb the bond matrix into the PEPS tensors connected by the bond. To this end, we make one more Schmidt decomposition
\be 
S~=~u~\mu~v^\dag~.
\label{umuvdag}
\ee
Here $u,v$ are unitary $D\times D$ matrices and $\mu$ is a diagonal matrix of singular values. In Fig. \ref{FigApp}C the matrix $u~\sqrt{\mu}$ is absorbed to the left PEPS tensor $B_1$, and the matrix $\sqrt{\mu}~v^\dag$ to the right tensor $B_2$, thus completing the renormalization procedure. Alternatively, the whole bond matrix $S$ can be simply absorbed to any of the two PEPS tensors connected by the bond. We use the more symmetric version to simulate evolution of PEPS in real time \cite{future}.


\begin{thebibliography}{99}

\bibitem{White} S. R. White, Phys. Rev. Lett. 69, 2863 (1992).

\bibitem{PEPS} F. Verstraete and J. I. Cirac, cond-mat/0407066; 
               V. Murg, F. Verstraete, and J. I. Cirac, 
               Phys. Rev. A 75, 033605 (2007);
               G. Sierra and M. A. Martın-Delgado, 
               arXiv:cond-mat/9811170; 
               T. Nishino and K. Okunishi, 
               J. Phys. Soc. Jpn 67 3066 (1998); 
               Y. Nishio, N. Maeshima, A. Gendiar, and T. Nishino, 
               cond-mat/0401115;
               J. Jordan, R. Or\'us, G. Vidal, F. Verstraete, and J. I. Cirac,
               Phys. Rev. Lett. 101, 250602 (2008);
               Z.-C. Gu, M. Levin, and X.-G. Wen, 
               Phys. Rev. B 78, 205116 (2008);
               H. C. Jiang, Z. Y. Weng, and T. Xiang, 
               Phys. Rev. Lett. 101, 090603 (2008);
               Z. Y. Xie, H. C. Jiang, Q. N. Chen, Z. Y. Weng, and T. Xiang, 
               Phys. Rev. Lett. 103, 160601 (2009);
               P.-C. Chen, C.-Y. Lai, and M.-F. Yang, 
               J. Stat. Mech.: Theory Exp. (2009) P10001;
               R. Or\'us and G. Vidal, 
               Phys. Rev. B 80, 094403 (2009).
               
\bibitem{MERA} G. Vidal, 
               Phys. Rev. Lett. 99, 220405 (2007); 
               G. Vidal,
               Phys. Rev. Lett. 101, 110501 (2008); 
               L. Cincio, J. Dziarmaga, and M. M. Rams, 
               Phys. Rev. Lett. 100, 240603 (2008); 
               G. Evenbly and G. Vidal, 
               Phys. Rev. Lett. 102, 180406 (2009); 
               G. Evenbly and G. Vidal, 
               Phys. Rev. B 79, 144108 (2009).

\bibitem{fermions}   P. Corboz, G. Evenbly, F. Verstraete, and G. Vidal, 
                     Phys. Rev. A 81, 010303(R) (2010);
                     C. V. Kraus, N. Schuch, F. Verstraete, and J. I. Cirac,
                     Phys. Rev. A 81, 052338 (2010);
                     C. Pineda, T. Barthel, and J. Eisert, 
                     Phys. Rev. A 81, 050303(R) (2010).

\bibitem{highTc} J. Hubbard, Proc. Roy. Soc. (London), Ser. A 276, 238 (1963); 
                 P. W. Anderson, Science 235, 1196 (1987).

\bibitem{tJ} 
	      P. Corboz, R. Or\'us, B. Bauer, and G. ́Vidal,
              Phys. Rev. B 81, 165104 (2010);
              P. Corboz, S. R. White, G. Vidal, and M. Troyer,
              Phys. Rev. B 84, 041108 (2011);
	      Q.-Q. Shi, S.-H. Li, J.-H. Zhao, H.-Q. Zhou,
	      arXiv:0907.5520;
              S.-H. Li, Q.-Q. Shi, H.-Q. Zhou,		 
	      arXiv:1001.3343.

\bibitem{VMC} D. A. Ivanov. Phys. Rev. B 70, 104503 (2004);
              W.-J. Hu, F. Becca, S. Sorella,
              Phys. Rev. B 85, 081110(R) (2012).

\bibitem{ancillas} F. Verstraete, J. J. Garcia-Ripoll, and J. I. Cirac,
                                             Phys. Rev. Lett. 93, 207204 (2004); 
                                             M. Zwolak and G. Vidal, 
                                             Phys. Rev. Lett. 93, 207205 (2004);
                                             A.E. Feiguin and S.R. White, 
                                  Phys. Rev. B 72, 220401 (2005).
                                  
\bibitem{WhiteT} S. R. White, Phys. Rev. Lett. 102, 190601 (2009);
                 E.M. Stoudenmire, and S. R. White,
                 New J. Phys. 12, 055026 (2010).        
                 
\bibitem{HOSVD} Z. Y. Xie, J. Chen, M. P. Qin, J. W. Zhu, L. P. Yang, and T. Xiang,
                Phys. Rev. B 86, 045139 (2012).                 
                     
\bibitem{PEPO} R. Or\'us,         
               Phys. Rev. B 85, 205117 (2012).           
                                        
\bibitem{CMR} R. J. Baxter, J. Math. Phys. 9, 650 (1968); J. Stat. Phys. 19, 461 (1978); 
              T. Nishino and K. Okunishi, J. Phys. Soc. Jpn. 65, 891 (1996);
              R. Or\'us and G. Vidal, Phys. Rev. B 80, 094403 (2009).

\bibitem{hc} H. Rieger, N. Kawashima, 
             Europ. Phys. J. B 9, 233 (1999);
             H.W.J. Blote and Y. Deng, 
             Phys. Rev. E 66, 066110 (2002).
             
\bibitem{future} P. Czarnik {\it et al.},
                 in preparation.             


\end{thebibliography}
\end{document}